\documentstyle[prc,aps,preprint]{revtex}
\begin{document}
\draft
\title{Collisional Damping of Nuclear Collective Vibrations in a Non-Markovian Transport Approach
\footnote{This work is supported in part by the US DOE Grant No. DE-FG05-89ER40530}}
\author{
S.\  Ayik$^{1},$ O. Yilmaz$^{2},$ A. Gokalp$^{2},$ P. Schuck$^{3},$   }
\address{$^{1}$Physics Department, Tennessee Technological University,
Cookeville, TN 38505, USA}
\address{$^{2}$Physics Department, Middle East Technical University,
06531 Ankara, Turkey}
\address{$^{3}$Institut des Sciences Nucleaires, 38026, Grenoble, France}
\date{\today}
\maketitle
\begin{abstract}
A detailed derivation of  the collisional widths of collective vibrations is
presented in both quantal and semi-classical frameworks by considering the
linearized limits of the extended TDHF and the BUU model with a non-Markovian
binary collision term. Damping widths of giant dipole and giant quadrupole excitations
are calculated by employing an effective Skyrme force, and the results are compared with
GDR measurements in Lead and Tin nuclei at finite temperature.
\end{abstract}

\narrowtext

\section{Introduction}

Excitation energy dependence of the giant resonance is still one of the 
open problems in the
study of collective modes in nuclei at finite temperatures. Recent experimental
investigations of the giant dipole resonance in the mass region 
$A$=110-140 \cite{Hofmann,Tin} and  the $^{208}Pb$ nucleus \cite{Lead} show that damping monotonically increases with excitation energy,
and in the former case, saturates at high excitations. 
In medium-weight and heavy nuclei at relatively low temperatures the
owerwhelming contribution to damping arises from the spreading width
$\Gamma^{\downarrow}$ due to mixing of collective states with 
more complicated states,
which is dominated by the coupling with 2p-2h doorway excitations
\cite{Wambach,BerBro,Lauritzen}. There are essentially two different 
approaches for calculation of the spreading width 
$\Gamma^{\downarrow}$: (i) Coherent 
mechanism due to coupling with low-lying surface modes which provides 
an important mechanism for damping of giant resonance in particular at 
low temperatures \cite{BerBorBro,Ormand}, (ii) Damping due to the coupling
with incoherent 2p-2h states which is usually referred to as the collisional
damping \cite{DiToro,Morawetz} and the Landau damping modified by 
two-body collisions \cite{Baran,Shlomo}. Calculations carried out on 
the basis of these approaches are partially successful in explaining 
the broadening of the giant dipole resonance with increasing temperature, 
but the saturation is still an open problem \cite{Alhassid}. In this work,
we do not consider the coherent contribution to the spreading width due to
the coupling with low-lying surface modes, but investigate in detail the 
collisional damping at finite temperature due to decay of the collective 
state into incoherent 2p-2h excitations in the basis of a non-Markovian
transport approach. 

Semi-classical transport models of the Boltzmann-Uehling-Uhlenbeck (BUU)
type are often employed for studying nuclear collective vibrations \cite{Bonasera}.
Although, these models give a good description for the average resonance
energies, they do not provide a realistic description for the collisional
relaxation rates. In these standard models, the collision term is treated
in a Markovian approximation by assuming that the two-body collisions are
local in both space and time, in accordance with Boltzmann's original
treatment. This simplification is usually justified by the fact that the
duration of two-body collisions is short on the time scale characteristic
of macroscopic evolution of the system. However, when the system possesses
fast collective modes with characteristic energies that are not small in
comparison with temperature, the standard Markovian treatment is inadequate.
It leads to an incorrect energy conservation factor in the collision
term, which severely restricts the available 2p-2h phase space for damping
of the collective modes. Therefore, for a proper description of the
collisional relaxation rates, it is necessary to improve the transport
model by including the memory effect due to finite duration of two-body
collisions \cite{Ando,AyiDwo,Reinhard}.

Recently, we have investigated the collisional damping of giant resonances 
within the linearized limit of 
the BUU model with a non-Markovian collision term, and derived closed form 
expressions for damping 
width of isoscalar and isovector collective vibrations \cite{AyiBoi,AyiBel}. 
Also, the model has been
applied to study the density fluctuations and the growth of instabilities in 
the nuclear matter within the 
stochastic Boltzmann-Langevin approach \cite{AyikZ,AyiRan}. As a result of the non-Markovian
structure of two-body collisions, in expressions of transport coefficients of collective modes 
(i.e. damping width and diffusion coefficient),  the available phase space for decay into 2p-2h states is 
properly taken into account with the correct energy conserving factor. In nuclear matter limit and for 
isotropic nucleon-nucleon cross-sections, by employing the standard approximation familiar in Fermi 
Liquid theory, it is possible to give analytical expression for the collisional width as
$\Gamma = \Gamma_{0}[(\hbar \omega)^2+(2\pi T)^2]$, where $\Gamma_{0}$ is different for different 
resonance and determined by nuclear matter properties and in-medium cross-sections. The quadratic 
temperature dependence fits well with the measured GDR widths in $^{120}Sn $ and $^{208}Pb $ 
nuclei, however the factor $\Gamma_{0}$ calculated with a cross-section of 40 mb underestimates the 
data by a factor of 2-3. In a recent work, we have performed calculations by employing energy and angle 
dependent free nucleon-nucleon cross-sections and by taking surface effects into account \cite{BelAyi}.

In this work, we give a brief description of the non-Markovian extension of the nuclear transport theory in 
both quantal and semi-classical frameworks, and present a detailed derivation of the collisional widths of 
collective vibrations. The derivation is carried out in both quantal and semi-classical frameworks by 
considering linearized limits of the extended TDHF and the BUU model with memory effects. The major 
uncertainty in the calculation of collisional widths arises from the lack of 
an accurate knowledge of the in-medium cross-sections in the vicinity of Fermi energy. The Skyrme force provides a good description of 
the imaginary part of the single particle optical potential and its radial dependence in the vicinity of Fermi 
energy \cite{SchHas}. Therefore, it may be suitable for describing in-medium effects in the collision term around Fermi 
energy. By employing an effective Skyrme force, we perform calculations of the damping widths of the 
giant dipole and giant quadruple excitations in semi-classical approximation and compare them with the 
GDR measurements in Lead and Tin nuclei at finite temperature.

\section{One-body transport model with memory effects}

In the extended TDHF approximation, the evolution of the single particle
density matrix $\rho(t)$ is determined by a transport equation \cite{Abe},
\begin{equation}
i\hbar \frac{\partial}{\partial t}\rho-[h,\rho]=K(\rho)
\end{equation}
where $h(\rho)$ is the mean-field Hamiltonian and the quantity on 
the right-hand-side represents a quantal collision term, which is specified by the 
correlated part of the two-particle density matrix as
$K(\rho_{1})= tr_{2}[v,C_{12}]$ with $v$ as the effective residual
interactions. The correlated part of the two-particle density matrix is defined as
$C_{12}=\rho_{12}-\widetilde{\rho_{1}\rho_{2}}$,
where $\widetilde{\rho_{1}\rho_{2}}$
represents the antisymmetrized product of the single-particle density
matrices, and it is determied 
by the second equation of the BBGKY hierarchy. Retaining only the
lowest order terms in the residual interactions, the hierarchy can be
truncated on the second level, and the hence the correlated part of the
two-particle density matrix evolves according to,
\begin{equation}
i\hbar \frac{\partial}{\partial t}C_{12}-[h,C_{12}]=F_{12}
\end{equation}
where
\begin{equation}
F_{12}=(1-\rho_{1})(1-\rho_{2})v\widetilde{\rho_{1}\rho_{2}}-
\widetilde{\rho_{1}\rho_{2}}v(1-\rho_{1})(1-\rho_{2}).
\end{equation}
An expression for the collision term can be obtained by formally
solving Eq.(2), and substituting the result into Eq.(1),
\begin{equation}
K(\rho)=-\frac{i}{\hbar}\int_{0}^{t} d\tau tr_{2}
[v, G(t,t-\tau)F_{12}(t-\tau)G^{\dagger}(t,t-\tau)]
\end{equation}
where
\begin{equation}
G(t,t-\tau)=T \cdot exp[-\frac{i}{\hbar}
\int_{t-\tau}^{t} dt' h(t')]
\end{equation}
denotes the mean-field propagator.

The transport equation (1) is usually considered in semi-classical
approximation. In this case, one deals with the phase-space density
$f({\bf r, p})$ defined as the Wigner transform of the density matrix,
\begin{equation}
f({\bf r, p})=\int \frac{d{\bf k}}{(2\pi\hbar)^3}
e^{-i\frac{{\bf k \cdot r}}{\hbar}}
<{\bf p}+\frac{{\bf k}}{2}|\rho|{\bf p}-\frac{{\bf k}}{2}>.
\end{equation}
Performing the Wigner transform of Eq.(1) and retaining the lowest order
terms in gradients in accordance with the standard treatment 
\cite{KadBay,Pawel},  yields a semi-classical transport equation for the 
phase-space density, 
\begin{equation}
\frac {\partial}{\partial t}f({\bf r, p}) -\{h(f), f({\bf r, p})\}=K(f).
\end{equation}
Here, the left hand side describes the Vlasov propagation in terms of the
self-consistent one-body Hamiltonian $h(f)$, and $K(f)$ represents the
collision term in semi-classical approximation, which has a non-Markovian
form due to the memory effects arising from the finite duration of two-body
collisions,
\begin{equation}
K(f)=\int d{\bf p}_2d{\bf p}_3d{\bf p}_4 \int_{0}^{t} d \tau
w(12;34;\tau) [(1-f) (1-f_2) f_3 f_4- f f_2(1-f_3) (1-f_4)]_{t-\tau}
\end{equation}
In this expression the phase-space density is evaluated at time $t-\tau$
according to
$f_j(t-\tau)=f({\bf r}-\tau {\bf p}_{j}/m,{\bf p}_{j}+\tau \nabla U,t-\tau)$
and the collision kernel is given by
\begin{equation}
w(12;34;\tau)=\frac{1}{2\pi}W(12;34)[ g_1(\tau) g_2(\tau) g_{3}^{*}(\tau) g_{4}^{*}(\tau)+c.c.]
\end{equation}
where  $ g_{j}(\tau)$ is the Wigner transform of the mean-field propagator
\begin{equation}
g_{j}(\tau )=\int \frac{d{\bf k}}{(2\pi\hbar)^3} e^{-i\frac{{\bf k \cdot r}}{\hbar}}
<{\bf  p}_{j}+\frac{{\bf k}}{2}|G(t, t- \tau)|{\bf  p}_{j}-\frac{{\bf k}}{2}>
\end{equation}
and $W(12;34)$ denotes the basic two-body transition rate
\begin{equation}
W(12;34) =  \frac {\pi}{(2\pi\hbar)^6} |<\frac{{\bf p}_{1}-{\bf p}_{2}}{2}|v
|\frac{{\bf p}_{3}-{\bf p}_{4}}{2}>_{A}|^{2}
\delta({\bf p}_{1}+{\bf p}_{2}-{\bf p}_{3}-{\bf p}_{4})
\end{equation}
which can be expressed in terms of the scattering cross-section as
\begin{equation}
W(12;34)= \frac{1}{(2\pi\hbar)^3}\frac{4\hbar}{m^2} \frac{d\sigma}{d\Omega}
\delta({\bf p}_{1}+{\bf p}_{2}-{\bf p}_{3}-{\bf p}_{4}).
\end{equation}

The collision term involves different characteristic time scales including
the average duration time of two-body collisions $\tau_{d}$, the
characteristic time associated with the mean-field fluctuations
$\tau_{mf}$ and the mean-free-time between two-body collisions
$\tau_{\lambda}$. The range of the integration over the past history
in the collision term (8)  (and also in (4)) is essentially determined by
the duration time
of two-body collisions. Usually, two-body collisions
are treated in a Markovian approximation by assuming  the duration time
of collisions is much shorter than the other time scales
$\tau_{d}\ll \tau_{mf},\tau_{\lambda}$, which would be appropriate if
two-body collisions can be considered instantaneous. In this case, the $\tau$
dependence of the phase-space density in the collision term can be neglected
$f_{j}(t-\tau) \approx  f_{j}(t)$, and the mean-field propagator can be
approximated by the kinetic energy alone,
$g_{j}\approx exp[-i \tau \epsilon_{j}]$ with $\epsilon_{j}=p_{j}^{2}/2m$.
This yields energy conserving two-body collisions, and the resultant
semi-classical transport equation is known as the BUU model \cite{BerGup}.
The standard description
provides a good approximation at intermediate energies when the system does
not involve fast collective modes, since the weak-coupling condition is well
satisfied due to relatively long mean-free-path of nucleons. When the system
possesses fast collective modes, for example high frequency collective vibrations
or rapidly growing unstable modes, the Markovian approximation breaks down
and  the influence of the mean-field fluctuations in the collision term becomes
important. The finite duration time allows for a direct coupling between two-body
collisions and the mean-field fluctuations, which strongly modifies the collisional
relaxation properties of the collective modes as compared to the Markovian
limit, in particular at low temperatures \cite{PhysKin}. In this work we consider the 
mean-field
dominated regime in which the nucleon mean-free-time is long as compared to
the characteristic time associated with the mean-field fluctuations and the duration time
of collisions
$\tau_{d}, \tau_{mf} \ll \tau_{\lambda}$,  which maybe referred as weakly non-Markovian regime.
In this case, the $\tau$ dependence of the phase-space density in the collision term
can be neglected, as before, and the collision term takes essentially
a Markovian form with an effective transition rate given by
$\int_{0}^{t} d \tau w(12;34;\tau)$. When all different time
scales are of the same order of magnitude, the collision term becomes strongly non-Markovian,
and the time evolution of the system is accompanied by off-shell two-body
collisions .

\section{Transport description of collective vibrations}

We apply the non-Markovian transport model developed in the previous section to describe
small amplitude collective vibrations around a stable equilibrium  in the
linear response approximation, and present an explicit derivation of the
expression for the collisional damping widths of the collective modes
in both quantal and semi-classical frameworks.

\subsection{Quantal treatment}

To describe small amplitude collective vibrations around a finite
temperature equilibrium state $\rho_{0}$, we linearize Eqs.(1) and (2)
for small deviations $\delta\rho=\rho-\rho_{0}$ and
$\delta C_{12}=C_{12}-C_{12}^{0}$,
\begin{equation}
i\hbar \frac{\partial}{\partial t}\delta\rho-
[\delta h,\rho_{0}]-[h_{0},\delta\rho]=tr_{2}[v,\delta C_{12}]
\end{equation}
and
\begin{equation}
i\hbar \frac{\partial}{\partial t}\delta C_{12}-
[\delta h,C_{12}^{0}]-[h_{0},\delta C_{12}]=\delta F_{12}
\end{equation}
where 
$\delta h=\left(\partial U/\partial \rho \right)_{0}\cdot \delta\rho$
represent the small deviations in the single-particle density matrix,
the quantity $\delta F_{12}$ is
\begin{eqnarray}
\delta F_{12} =
 -\delta \rho_{1}(1-\rho_{2}^{0})v\widetilde{\rho_{1}^{0}\rho_{2}^{0}}
 -(1-\rho_{1}^{0})\delta \rho_{2}v\widetilde{\rho_{1}^{0}\rho_{2}^{0}}+
\nonumber \\
(1-\rho_{1}^{0})(1-\rho_{2}^{0})v\widetilde{\delta\rho_{1}\rho_{2}^{0}}
+(1-\rho_{1}^{0})(1-\rho_{2}^{0})v\widetilde{\rho_{1}^{0}\delta\rho_{2}}-h.c.
\end{eqnarray}
and the equilibrium correlation function $C_{12}^{0}$ is determined by
\begin{equation}
-[h_{0},C_{12}^{0}]=F_{12}^{0}
\end{equation}
with $F_{12}^{0}$ as the equilibrium value of $F_{12}$.

We can analyze the collective vibration by expanding the small deviations
$\delta \rho$ in terms of normal modes of the system \cite{Ring},
\begin{equation}
\delta \rho(t)=\sum[z_{\lambda}(t) \rho_{\lambda}^{\dagger}+ z_{\lambda}^{*}(t) \rho_{\lambda}]
\end{equation}
where $\rho_{\lambda}^{\dagger}$ and $\rho_{\lambda}$ represent the normal modes
of the system. When the damping width is small as compared to the mean frequency of the
mode, we can follow a perturbation approach and determine the normal modes 
by the standard RPA equations without including the collision term,
\begin{equation}
\hbar\omega_{\lambda}\rho_{\lambda}^{\dagger}-[h_{\lambda}^{\dagger},\rho_{0}]
-[h_{0},\rho_{\lambda}^{\dagger}]=0.
\end{equation}
Here, $\omega_{\lambda}$ is the frequency of the normal mode and
$h_{\lambda}^{\dagger}$ represents the positive frequency part of the vibrating
mean-field. It is convenient to introduce the RPA amplitudes
$\hat{O}_{\lambda}^{\dagger}$
and $\hat{O}_{\lambda}$ associated with normal modes according to
$\rho_{\lambda}^{\dagger}=[\hat{O}_{\lambda}^{\dagger},\rho_{0}]$
and its hermitian conjugate.
In the representation which diagonalizes $\rho_{0}$, the RPA amplitudes can be expressed as
\begin{equation}
<n|\hat{O}_{\lambda}^{\dagger}|m>=\frac{<n|h_{\lambda}^{\dagger}|m>}
{\hbar \omega_{\lambda}-\epsilon_{n }+\epsilon_{m}}
\end{equation}
and they are normalized as
$tr[\hat{O}_{\lambda},\hat{O}_{\lambda}^{\dagger}]\rho_{0}=1$.
Substituting the expansion (17)
into Eq.(13) and projecting by  $\hat{O}_{\lambda}$ yields
\begin{equation}
\frac{dz_{\lambda}}{dt}+i\omega_{\lambda} z_{\lambda}=
-\frac{1}{2}\Gamma_{\lambda}z_{\lambda}
\end{equation}
for the amplitudes of the normal modes. These amplitudes execute
a damped harmonic motion with a damping coefficient given by
\begin{equation}
\Gamma_{\lambda}=tr[\hat{O}_{\lambda},v]C_{\lambda}^{\dagger}
\end{equation}
and it describes the spreading width of the RPA mode due to coupling with
the two particle-two hole states, which is usually referred to as
the collisional damping. In this expression, $C_{\lambda}^{\dagger}$ denotes
the positive frequency part of the correlations  and it is determined by
\begin{equation}
\hbar \omega_{\lambda}C_{\lambda}^{\dagger}-[h_{\lambda}^{\dagger},C_{12}^{0}]
-[h_{0},C_{\lambda}^{\dagger}]=F_{\lambda}^{\dagger}
\end{equation}
where $F_{\lambda}^{\dagger}$ represents the positive frequency part of
$\delta F_{12}$. In the representation diagonalizing $\rho_{0}$,
the correlation can be expressed as
\begin{eqnarray}
<nm|C_{\lambda}^{\dagger}|kl>=
\frac{<nm|[h_{\lambda}^{\dagger},C_{12}^{0}]+F_{\lambda}^{\dagger}|kl>}
{\hbar \omega_{\lambda}-\epsilon_{n}- \epsilon_{m}+ \epsilon_{k}+
\epsilon_{l}-i\eta} \nonumber \\
=\frac{<nm|[\hat{O}_{\lambda}^{\dagger},F_{12}^{0}]+F_{\lambda}^{\dagger}|kl>}
{\hbar \omega_{\lambda}-\epsilon_{n}- \epsilon_{m}+ \epsilon_{k}+
\epsilon_{l}-i\eta}.
\end{eqnarray}
According to expression (16), the matrix elements of the equilibrium correlation is
given by $<nm|C_{12}^{0}|kl >=<nm|F_{12}^{0}|kl>/
(\epsilon_{k}+ \epsilon_{l}- \epsilon_{n}- \epsilon_{m}-i\eta)$, in which
only the principal value part is non-vanishing. The second line of the above expression is obtained
by replacing the energy factors in the intermediate states according to
$\delta(\hbar \omega_{\lambda}-\epsilon_{n}- \epsilon_{m}+ \epsilon_{k}+
\epsilon_{l})$ and using the
definition of the RPA amplitudes. Furthermore following the observation
\begin{equation}
[\hat{O}_{\lambda}^{\dagger },F_{12}^{0}]= -F_{\lambda}^{\dagger}+
(1-\rho_{1}^{0})(1-\rho_{2}^{0}) [\hat{O}_{\lambda}^{\dagger},v]
\widetilde{\rho_{1}^{0}\rho_{2}^{0 }}-
\widetilde{\rho_{1}^{0}\rho_{2}^{0}}[\hat{O}_{\lambda}^{\dagger},v]
(1-\rho_{1}^{0})(1-\rho_{2}^{0})
\end{equation}
the correlation can be expressed as,
\begin{equation}
<nm|C_{\lambda}^{\dagger}|kl>=<nm|[\hat{O}_{\lambda}^{\dagger},v]|kl>_{A}
\frac{[\rho_{n}\rho_{m}\bar{\rho}_{k}\bar{\rho}_{l}-
\rho_{k}\rho_{l}\bar{\rho}_{n}\bar{\rho}_{m}]}{\hbar \omega_{\lambda}-
\epsilon_{n}- \epsilon_{m}+ \epsilon_{k}+ \epsilon_{l}-i\eta}
\end{equation}
where $\rho_{n}$ denotes the Fermi-Dirac occupation factor,
$\bar{\rho}_{n}=1-\rho_{n}$ and
$<nm|[\hat{O}_{\lambda}^{\dagger},v]|kl>_{A}$
represents the anti-symmetric matrix elements.
As a result, the damping width of the RPA states is given by \cite{AyiDwo},
\begin{eqnarray}
\Gamma_{\lambda}& = &\frac{\pi}{2}  \sum 
|<nm|[\hat{O}_{\lambda},v]|kl>_{A}|^{2}
\delta(\hbar \omega_{\lambda}-\epsilon_{n}- \epsilon_{m}+ \epsilon_{k}+
\epsilon_{l}) \nonumber  \\
&  & [\rho_{k}\rho_{l}\bar{\rho}_{n}\bar{\rho}_{m}-
\rho_{n}\rho_{m}\bar{\rho}_{k}\bar{\rho}_{l}].
\end{eqnarray}
The same expression for the damping width has been derived 
in ref.\cite{Mori} by employing a different approach.
It also can be obtained  in using the Green's function method
\cite{Adachi}, or more intuitive approaches \cite{BerBorBro}. It should
 be mentioned that the expression (26) was written down for the
first time by Landau, and it has become  a classical result of 
Fermi Liquid theory. However in the nuclear physcis literature several
expressions of (26) exist  which are at  variance with Landau's result \cite{Landau}.
The subtlety hinges on the expression (19) for $\hat{O}_{\lambda}^{\dagger}$.
Indeed, contrary to the ordinary RPA amplitudes, which at zero temperature have only
ph or hp components, no phase space factors appear
in (19), thus allowing non-zero values of 
$<n|\hat{O}_{\lambda}^{\dagger}|m>$ also for pp and hh
configurations (see also expression (30), below).

\subsection{Semi-classical treatment}

It is possible to describe the collective vibrations in semi-classical
approximation. In this case, one considers the phase-space density
$\delta f({\bf r, p})$ associated with small amplitude vibrations.
The equation of motion of the small amplitude vibrations in the semi-classical limit
is obtained by linearizing the transport equation (7),
\begin{equation}
\frac {\partial}{\partial t} \delta f({\bf r, p})+ {\bf v}\cdot \nabla \delta f({\bf r, p})-
{\bf v}\cdot \nabla \delta h \frac{\partial}{\partial \epsilon} f_{0} = \delta K({\bf r,p})
\end{equation}
where $\delta K({\bf r,p})$ denotes the linearized collision term,
\begin{equation}
\delta K({\bf r,p}) =- \frac{i}{\hbar}\int \frac {d{\bf k}}{(2\pi\hbar)^3}
e^{-i\frac{{\bf k \cdot r}}{\hbar}}d{\bf p}_{2}
<{\bf p}+\frac{{\bf k}}{2},{\bf p}_{2}|[v,\delta C_{12}]
|{\bf p}-\frac{{\bf k}}{2},{\bf p}_{2}>
\end{equation}
${\bf v}={\bf p}/m$ and the equilibrium state $f(\epsilon)$ is taking to
be homogeneous for simplicity.
In a manner similar to quantal treatment, the phase-space density
can be expanded in terms of normal modes as
\begin{equation}
\delta f({\bf r,p})=\sum[-i\hbar z_{\lambda}(t) {\bf v} \cdot \nabla
O_{\lambda}^{*}+i\hbar z_{\lambda}^{*} {\bf v} \cdot \nabla
O_{\lambda}] \frac{\partial}{\partial \epsilon}f
\end{equation}
where $O_{\lambda}^{*}$ and $O_{\lambda}$ are the
Wigner transform of the RPA amplitudes $\hat{O}_{\lambda}^{\dagger}$
and $\hat{O}_{\lambda}$. In the perturbation approach, these amplitudes are
 given by
\begin{equation}
O_{\lambda}^{*}({\bf r,p})=
(\frac{1}{\hbar \omega-i\hbar {\bf v} \cdot \nabla})\cdot
h_{\lambda}({\bf r})
\end{equation}
and its complex conjugate. In a similar manner, we can  expand the
correlation function in terms of the normal modes as
$\delta C_{12}(t)=\sum[z_{\lambda}(t) C_{\lambda}^{\dagger}+
z_{\lambda}^{*}(t) C_{\lambda}]$. By inserting this expansion
 into Eq.(28),  we obtain an expression of the collision term
in terms of the RPA amplitudes,
\begin{eqnarray}
\delta K({\bf r,p}) & = & \frac {1}{2}  \sum_{\lambda}
\int  d{\bf p}_{2} d{\bf p}_{3} d{\bf p}_{4} W(12;34)
[z_{\lambda}\Delta O_{\lambda}^{*}+z_{\lambda}^{*}\Delta O_{\lambda}] \\
& &[\delta(\hbar\omega_{\lambda}-\Delta\epsilon)-
\delta(\hbar\omega_{\lambda}+\Delta\epsilon)]
[\bar{f}_{1} \bar{f}_{2}f_{3}f_{4}-
f_{1}f_{2}{\bar f}_{3}{\bar f}_{4}] \nonumber
\end{eqnarray}
where,
$\Delta \epsilon=\epsilon_{3}+\epsilon_{4}-\epsilon_{1}-\epsilon_{2}$,
$\Delta O_{\lambda}=O_{\lambda}(3)+
O_{\lambda}(4)-O_{\lambda}(1)-O_{\lambda}(2)$ with $\epsilon(j)=p_{j}^{2}/2m$
and $O_{\lambda}(j)=O_{\lambda}({\bf r,p_{j}})$, and $W(12;34)$ is the
transition rate given by eqs.(11) or (12). Substituting the normal mode decomposition of the phase-space 
density into (27) and carrying out a projection with
$O_{\lambda}$, we find the expression
\begin{eqnarray}
\Gamma_{\lambda} &=& \frac {1}{2} \frac {1}{(2\pi\hbar)^3} \int
d{\bf r} d{\bf p}_{1} d{\bf p}_{2} d{\bf p}_{3} d{\bf p}_{4}
|\Delta O_{\lambda}|^{2} W(12;34)   \\
& & [\delta(\hbar\omega_{\lambda}-\Delta\epsilon)-\delta(\hbar\omega_{\lambda}+
\Delta\epsilon)] \bar{f}_{1} \bar{f}_{2}f_{3}f_{4} \nonumber
\end{eqnarray}
for the collisional width, where the normal modes in the semi-classical approximation are normalized 
according to
\begin{equation}
-\int \frac {i}{(2\pi\hbar)^{3}} d{\bf r} d{\bf p} O_{\lambda}
{\bf v} \cdot \nabla O_{\lambda}^{*} \frac{\partial}{\partial \epsilon}f=1.
\end{equation}
We note that, this result for the collisional width can directly be
obtained by evaluating the quantal expression (26) in the
Thomas-Fermi approximation \cite{ThoFer}. We also note that in order to obtain the expressions
(26)  and (32) for the collisional width  in quantal  or semi-classical forms, the non-Markovian collision 
term should be linearized in a consistent manner by including the
contributions arising from the mean-field propagator and the phase-space factors. The 
result is, then, consistent with the Landau's expression for damping of zero sound modes,
and also, is in accordance with the quantal fluctuation-dissipation relation \cite{AyikZ,AyiRan}.
If the term involving the mean-field fluctuations (the second term in the
left-hand-side of eq.(14)) is ignored, one obtains a wrong expression for the 
collisional damping which gives a value that is factor of three larger than
its correct value in the nuclear matter \cite{AyiBoi}.
   
It is more convenient to express the semi-classical RPA modes in terms
of real functions $Q_{\lambda}({\bf r,p})$ and $P_{\lambda}({\bf r,p})$
defined as
\begin{equation}
Q_{\lambda}({\bf r,p})=
\frac{1}{\sqrt{2\omega_{\lambda}}}[O_{\lambda}^{*}({\bf r,p})+
O_{\lambda}({\bf r,p})]
\end{equation}
and
\begin{equation}
P_{\lambda}({\bf r,p})=
i\sqrt{\frac{\omega_{\lambda}}{2}}[O_{\lambda}^{*}({\bf r,p})-
O_{\lambda}({\bf r,p})].
\end{equation}
As a result, the normal mode expansion (29) becomes,
\begin{equation}
\delta f({\bf r,p})=\sum[q_{\lambda}(t)\chi_{\lambda}^{q}({\bf r,p})+
p_{\lambda}(t)\chi_{\lambda}^{p}({\bf r,p})]
(-\frac{\partial}{\partial \epsilon}f)
\end{equation}
where $\chi_{\lambda}^{q}=-\hbar {\bf v} \cdot \nabla P_{\lambda}$ and
$\chi_{\lambda}^{p}=-\hbar {\bf v} \cdot \nabla Q_{\lambda}$
represent the distortion factors of the phase-space density
associated with the real variables
$q_{\lambda}=\frac{1}{\sqrt{2\omega_{\lambda}}}[z_{\lambda}^{*}+z_{\lambda}]$
and
$p_{\lambda}=i\sqrt{\frac{\omega_{\lambda}}{2}}[z_{\lambda}^{*}-z_{\lambda}]$,
respectively. In the collision term (31), the factor
$z_{\lambda}\Delta O_{\lambda}^{*}+z_{\lambda}^{*}\Delta O_{\lambda}$
is replaced by
\begin{eqnarray}
z_{\lambda}\Delta O_{\lambda}^{*}+z_{\lambda}^{*}\Delta O_{\lambda}&=&
i\frac{z_{\lambda}}{\omega_{\lambda}}({\bf v} \cdot \nabla)
\Delta O_{\lambda}^{*}-i\frac{z_{\lambda}^{*}}{\omega_{\lambda}}
({\bf v} \cdot \nabla) \Delta O_{\lambda}\\
&=& \frac{1}{\hbar \omega_{\lambda}}
[q_{\lambda}\chi_{\lambda}^{q}({\bf r,p})+
p_{\lambda}\chi_{\lambda}^{p}({\bf r,p})]\nonumber
\end{eqnarray}
where the first line follows from an
identity satisfied by the semi-classical RPA amplitudes,
$O_{\lambda}^{*}({\bf r,p})=[h_{\lambda}({\bf r})+
i\hbar {\bf v} \cdot \nabla O_{\lambda}^{*}({\bf r})]/\hbar \omega_{\lambda}$.
In order to deduce the equations for the real variables $q_{\lambda}(t)$ and
$p_{\lambda}(t)$, we substitute the expansion (36) into Eq.(27) and
project the resultant equation by $O_{\lambda}$ and $P_{\lambda}$,
or equivalently by $\chi_{\lambda}^{q}$ and $\chi_{\lambda}^{p}$.
This gives two coupled equations for $q_{\lambda}(t)$ and $p_{\lambda}(t)$,
which can be combined to yield an equation in the form of a damped harmonic
oscillator, 
\begin{equation}
\ddot{q}_{\lambda}+
\left(\omega_{\lambda}^{2}+(\frac {\Gamma_{\lambda}}{2 \hbar})^2 \right) q_{\lambda}=
-\frac{\Gamma_{\lambda}}{\hbar} \dot {q}_{\lambda}
\end{equation}
where the collisional width is given by
\begin{eqnarray}
\Gamma_{\lambda}&=&\frac {1}{(2\pi\hbar)^3}  \int
d{\bf r} d{\bf p}_{1} d{\bf p}_{2} d{\bf p}_{3} d{\bf p}_{4}
[(\Delta \chi_{\lambda}^{q})^{2}+(\Delta \chi_{\lambda}^{p})^{2}]
W(12;34) \\
& & [\frac{\delta(\hbar\omega_{\lambda}-\Delta\epsilon)-
\delta(\hbar\omega_{\lambda}+\Delta\epsilon)}{4 \hbar \omega_{\lambda}}]
\bar{f}_{1} \bar{f}_{2}f_{3}f_{4} \nonumber
\end{eqnarray}
with the distortion factors normalized according to
\begin{equation}
\int \frac {1}{(2\pi\hbar)^{3}} d{\bf r} d{\bf p}
(\chi_{\lambda}^{q})^{2} (-\frac{\partial}{\partial \epsilon}f)=
\int \frac {1}{(2\pi\hbar)^{3}} d{\bf r} d{\bf p}
(\chi_{\lambda}^{p})^{2} (-\frac{\partial}{\partial \epsilon}f)=1.
\end{equation}
This expression, which is equivalent to the one given by Eq.(32),
provides a useful formula to calculate collisional damping
in terms of distortion factors of the momentum distribution
associated with the collective modes. The distortion factors may
be determined from the RPA treatment, or can be directly parametrized on
physical grounds. In practice, only one of the factors,
$\chi_{\lambda}^{q}$ or $\chi_{\lambda}^{p}$ which is associated
with a distortion of the momentum distribution, contributes the
collisional damping.

Spin-isospin effects in collective vibration can be easily incorporated
in the semi-classical RPA treatment by considering proton and neutron
degrees of freedom separately. The small deviations of the phase-space
densities $\delta f_{p}({\bf r,p})$, $\delta f_{n}({\bf r,p})$ of
protons and neutrons are determined by two coupled equations analogous
to Eq.(27). The collision terms in these equations involve binary collisions
between proton-proton, neutron-neutron and proton-neutron, 
and a summation over
the spins of the colliding particles. Observing that in isoscalar/isovector
modes protons and neutrons vibrate in-phase/out-of phase,
$\delta f_{p}({\bf r,p})=\mp\delta f_{n}({\bf r,p})$,
we can deduce equation of motions
for describing isoscalar/isovector vibrations by adding and subtracting
the corresponding equations for protons and neutrons. Carrying out the
semi-classical RPA treatment presented above, we obtain
$\Gamma_{\lambda}=\int d{\bf r}\Gamma_{\lambda}(r)$ with
\begin{equation}
\Gamma_{\lambda}^{s}(r)= \frac {1}{N_{\lambda}}
\int d{\bf p}_{1} d{\bf p}_{2} d{\bf p}_{3} d{\bf p}_{4}
[W_{pp}+W_{nn}+2W_{pn}] \left ( \frac {\Delta \chi_{\lambda}}{2}\right )^{2}Z 
f_{1}f_{2}\bar{f}_{3}\bar{f}_{4}
\end{equation}
and
\begin{equation}
\Gamma_{\lambda}^{v}= \frac {1}{N_{\lambda}}
\int d{\bf p}_{1} d{\bf p}_{2} d{\bf p}_{3} d{\bf p}_{4}
[(W_{pp}+W_{nn}) \left ( \frac {\Delta \chi_{\lambda}}{2}\right )^{2}+
2W_{pn}\left ( \frac {\widetilde {\Delta \chi}_{\lambda}}{2} \right )^{2}] Z
f_{1}f_{2}\bar{f}_{3}\bar{f}_{4}
\end{equation}
for the collisional widths of isoscalar and isovector modes, respectively.
Here, $N_{\lambda}=\int d{\bf r} d{\bf p}(\chi_{\lambda})^{2}
(-\frac{\partial}{\partial \epsilon}f)$ is a normalization,
$\Delta \chi_{\lambda}=\chi_{\lambda}(1)+\chi_{\lambda}(2)-
\chi_{\lambda}(3)-\chi_{\lambda}(4)$,
$\widetilde{\Delta \chi}_{\lambda}=\chi_{\lambda}(1)-\chi_{\lambda}(2)-
\chi_{\lambda}(3)+\chi_{\lambda}(4)$, and
$Z= [\delta(\hbar\omega_{\lambda}-\Delta\epsilon)-
\delta(\hbar\omega_{\lambda}+\Delta\epsilon)]/ \hbar\omega_{\lambda}$.
In these expressions, transition rates associated with
proton-proton, neutron-neutron and proton-neutron collisions are given by
eq.(12) with the corresponding cross-sections
\begin{eqnarray}
\lefteqn{\left(\frac{d\sigma}{d\Omega}\right)_{pp}=
\left(\frac{d\sigma}{d\Omega}\right)_{nn}=}\\
& &\frac{\pi}{(2\pi\hbar)^3}\frac{m^2}{4\hbar}\cdot
\frac{1}{4}\sum_{S} (2S+1)|<\frac{{\bf p}_{1}-{\bf p}_{2}}{2};S,T=1|v
|\frac{{\bf p}_{3}-{\bf p}_{4}}{2};S,T=1>_{A}|^{2} \nonumber
\end{eqnarray}
and
\begin{equation}
\left(\frac{d\sigma}{d\Omega}\right)_{pn}=
\frac{\pi}{(2\pi\hbar)^3}\frac{m^2}{4\hbar}\cdot
\frac{1}{8}\sum_{S,T} (2S+1)|<\frac{{\bf p}_{1}-{\bf p}_{2}}{2};S,T|v
|\frac{{\bf p}_{3}-{\bf p}_{4}}{2};S,T>_{A}|^{2}
\end{equation}
where $<{\bf p}_{1}-{\bf p}_{2}/2;S,T|v|{\bf p}_{3}-{\bf p}_{4}/2;S,T>_{A}$
represents the
fully antisymmetric two body matrix element of the residual
interaction between states with total spin and isospin $S$ and $T$.
The residual interactions $v$ should be understood as an effective density
dependent force. It can indeed be shown that a reasonable approximation
for $v$ is the so-called $G$-matrix \cite{Ring}. Microscopic $G$-matrices
are not very practical for explicit use, and thus we adopt below a more 
phenomenological point of view replacing the $G$-matrix by one of the more
recent Skyrme forces. In doing so, we, however, should be carefull, since
in the vicinity of nuclear surface Skyrme-type forces usually do not
match at all to the free nucleon-nucleon cross-sections.

\section{Damping of GD and GQ excitations}

We apply the formulas (41) and (42) to calculate the collisional widths
of the giant quadruple and dipole modes by parametrizing the distortion
factors of the momentum distribution in terms of Legendre functions as
$\chi_{Q}=p^2 P_{2}(cos \theta)$ and $\chi_{D}=p P_{1}(cos \theta)$.
In our calculations, we employ an effective Skyrme force, which is
parameterized as
\begin{eqnarray}
v & = & t_{0}(1+x_{0}P_{\sigma})\delta({\bf r_{1}-r_{2}})+\frac{t_{1}}{2}
[\delta({\bf r_{1}-r_{2}}) \hat{{\bf k}}^2 +\hat {{\bf k}}\prime^2 \delta({\bf r_{1}-r_{2}})]+\nonumber 
\\
& & t_{2} \hat {{\bf k}}\prime \cdot \delta({\bf r_{1}-r_{2}}) \hat {{\bf k}}+
\frac{t_{3}}{6} \rho^{\alpha}\delta({\bf r_{1}-r_{2}})
\end{eqnarray}
where $\hat {{\bf k}}=({\bf p}_{1}-{\bf p}_{2})/2\hbar$ represents the relative
momentum operator with $\hat {{\bf k}}$ is acting to right and $\hat{{\bf k}}\prime$ is
acting to left. In the case of the quadruple mode the
collisional width is determined by the spin-isospin averaged nucleon-nucleon
cross-section, $(d\sigma/d\Omega)_{0}=[(d\sigma/d\Omega)_{pp}
+(d\sigma/d\Omega)_{nn}+2(d\sigma/d\Omega)_{pn}]/4$. In the case of
the dipole mode the only contribution comes from the spin averaged proton-neutron
cross-section, $(d\sigma/d\Omega)_{pn}$. In terms of the effective
Skyrme force these cross-sections are given by
\begin{eqnarray}
\left(\frac{d\sigma}{d\Omega}\right)_{0}  =  \frac{\pi}{(2\pi\hbar)^3}
\frac{{m^{\ast}} ^2}{4\hbar}
&&\left(\frac{3}{4}[t_{0}(1-x_{0})+\frac{t_{1}}{2}({\bf k}^2+{\bf k}\prime^2)+
\frac{t_{3}}{6}\rho^{\alpha}]^2+\frac{5}{2} [t_{2}{\bf k \cdot k\prime}]^2+\right.
\nonumber \\
& &\left.\frac{3}{4}[t_{0}(1+x_{0})+\frac{t_{1}}{2}({\bf k}^2+{\bf k}\prime^2)+
\frac{t_{3}}{6}\rho^{\alpha}]^2 \right)
\end{eqnarray}
and
\begin{eqnarray}
\left(\frac{d\sigma}{d\Omega}\right)_{pn}  =  \frac{\pi}{(2\pi\hbar)^3}
\frac{{m^{\ast}}^2}{4\hbar}
&&\left(\frac{1}{2}[t_{0}(1-x_{0})+\frac{t_{1}}{2}({\bf k}^2+{\bf k}\prime^2)+
\frac{t_{3}}{6}\rho^{\alpha}]^2+2 [t_{2}{\bf k \cdot k\prime}]^2+\right.
\nonumber \\
& &\left.\frac{3}{2}[t_{0}(1+x_{0})+\frac{t_{1}}{2}({\bf k}^2+{\bf k}\prime^2)+
\frac{t_{3}}{6}\rho^{\alpha}]^2 \right)
\end{eqnarray}
where ${\bf k}=({\bf p}_{1}-{\bf p}_{2})/2\hbar$ and
${\bf k\prime}=({\bf p}_{3}-{\bf p}_{4})/2\hbar$ are the relative momenta before
and after the binary colision, and $m^{\ast}$ denotes the effective mass
\begin{equation}
\frac{1}{m^{\ast}(r)}=\frac{1}{m}[1+\frac{2m}{\hbar^2}\frac{1}{16}(3t_1+5t_2)\rho(r)].
\end{equation}

In the bulk of the nucleus the Pauli blocking is very effective, and hence, the
overwhelming contributions to momentum integrals in expressions (41) and (42)
arise in the vicinity of the Fermi surface.  We can approximately perform these integrals
by employing the standard coordinate transformation \cite{Abrikosov},
\begin{equation}
\int d{\bf p}_{1} d{\bf p}_{2} d{\bf p}_{3} d{\bf p}_{4}
\delta({\bf p}_{1}+{\bf p}_{2}-{\bf p}_{3}-{\bf p}_{4}) \cdots \approx
\int p_{F} \frac{{m^{\ast}} ^4}{2}
d\epsilon_{1}d\epsilon_{2}d\epsilon_{3}d\epsilon_{4}
\frac{d\Omega_{1} d\Omega}{cos\theta/2} d\phi_{2} \cdots
\end{equation}
Furthermore, for temperatures small compared to the Fermi energy,
$T\ll \epsilon_{F}$, the energy integrals can be calculated 
analytically using the formula \cite{Abrikosov,BayPet},
\begin{equation}
\int d\epsilon_{1}d\epsilon_{2}d\epsilon_{3}d\epsilon_{4}
\delta(\hbar\omega\pm\Delta\epsilon)
\bar{f_{1}}\bar{f_{2}}f_{3}f_{4}\approx \mp \frac{\hbar\omega}{6}
\frac{(\hbar\omega)^2+(2\pi T)^2}{1-\exp(-\hbar\omega/T)}.
\end{equation}
Then, the bulk contribution to the collisional widths of the
quadrupole and dipole modes can be expressed as,
\begin{equation}
\Gamma_{Q}^{bulk}(r)=\frac {\hbar}{N_{Q}}\frac {4 \pi}{5} {m^{\ast}} ^2 \rho p_{F}^2
I_{Q}(r) [(\hbar\omega)^2+(2\pi T)^2]
\end{equation}
and
\begin{equation}
\Gamma_{D}^{bulk}(r)=\frac {\hbar}{N_{D}}\frac {2\pi}{3} {m^{\ast}}^2 \rho
I_{D}(r)[(\hbar\omega)^2+(2\pi T)^2].
\end{equation}
Here, $\rho$ denotes the particle density,
$\rho=(4/(2\pi\hbar)^3)(4\pi/3)p_{F}^3$,
the normalizations are $N_{Q}=(4\pi/5) \int d {\bf r}m^{\ast} p_{F}^5$,
$N_{D}=(4\pi/3) \int d {\bf r}m^{\ast} p_{F}^3$,
and the quantities $I_{Q}$, $I_{D}$ are given by
\begin{equation}
I_{Q}=\int \sin\frac{\theta}{2} d\theta d\phi
[1+P_{2}(\cos\theta)-2P_{2}(\cos\theta'_{3})]
\left(\frac{d\sigma}{d\Omega}\right)_{0}
\end{equation}
and
\begin{equation}
I_{D}=\int \sin\frac{\theta}{2} d\theta d\phi
[1-P_{1}(\cos\theta)] \left(\frac{d\sigma}{d\Omega}\right)_{pn}.
\end{equation}
In these expressions, the angular integrals can be performed analytically by noting that in the vicinity of 
Fermi surface the momentum dependent terms in the cross-sections
can be expressed in terms of  the standard  variables as
${\bf k}\cdot {\bf k}\prime=-k_{F}^2 \sin^{2}\theta/2\cos\phi$
and ${\bf k}^2={\bf k}\prime^2=k_{F}^2 \sin^{2}\theta/2$,
and
$\cos\theta'_{3}=(\cos\theta/2)^2-(\sin\theta/2)^2\cos\phi$.
As already mentioned earlier, the cross-sections based on Skyrme
forces have a strongly erroneous behaviour at very low densities.
For this reason, we can not use expressions (51) and (52) far out
in the surface. It is therefore absolutely
necessary to develope effective forces which have the correct
free cross-section limit. For the time being, we develope an interpolation
scheme.
In the vicinity of nuclear surface, $\rho(r) \ll \rho_{0}$, the Pauli
blocking is not effective. In this case, it is convenient to transform 
the integration variables
in (41) and (42) into the total momenta
${\bf  P}={\bf p}_1+{\bf p}_2$, ${\bf  P}\prime ={\bf p}_3+{\bf p}_4$ and relative
momenta ${\bf q}=({\bf p}_1-{\bf  p}_2)/2$,
${\bf q}\prime=({\bf p}_3-{\bf p}_4)/2$ before and after the collision.
Due to the energy conservation, the magnitude of the relative momentum after
the collision is restricted according to $q\prime= \sqrt {q^2 \mp m^{\ast} \hbar \omega} 
$.
In the tail region for $\epsilon_{F}, T \ll \hbar \omega $,
the expressions (41) and (42) may be estimated by omitting the Pauli blocking factors and
neglecting the $q$-dependent terms. This gives,
\begin{equation}
\Gamma_{Q}^{surf}(r) \approx \frac {\hbar}{N_{Q}} \frac{2\pi}{3}\rho p_{F}^3
(m^{\ast} \hbar \omega)^{3/2}I_{Q}(r)
\end{equation}
and
\begin{equation}
\Gamma_{D}^{surf}(r) \approx \frac {\hbar}{N_{D}} \frac{\pi}{3}\rho
p_{F}^3 (m^{\ast} \hbar \omega)^{1/2}I_{D}(r).
\end{equation}
We define an effective Pauli blocking factor as the ratio of the damping width with and without the Pauli 
blocking factors in expressions (41) and (42), 
$F_{\lambda}= \Gamma_{\lambda}(r)/ 
\Gamma_{\lambda}^{no Pauli}(r)$, for $\lambda = Q$ or $D$, and parameterize it in the following form,
\begin{equation}
F_{\lambda}(r)=1+\left (\frac{\epsilon_{F}(r)}{\epsilon_{F}(0)}\right )^{\beta}
\left ( F_{\lambda}(0)-1 \right )
\end{equation}
where  $\epsilon_{F}(0)$ is the Fermi energy at the bulk corresponding to the central density $\rho_{0}$, 
$ F_{\lambda}(0)$ is the effective factor at the bulk and 
$\beta = \hbar \omega_{\lambda}/2\epsilon_{F}(0)$. As a function of $r$, the effective blocking factor 
remains essentially constant and equals to its bulk value until the density reaches about $1/2$ of the 
central density, and then it smoothly goes to one at the surface of the nucleus. This form provides a good 
approximation for the exact calculations of the effective Pauli blocking of $2p-1h$ excitations in 
connection with the collisional damping of single particle states as reported in \cite{Hasse}.  Therefore we 
expect, it provides a reasonable approximation for $2p-2h$ excitations, and  calculate the collisional 
widths in an approximate manner by smoothly joining the bulk contribution with the surface contribution 
in accordance with the approximate blocking factor (57), 
\begin{equation}
\Gamma_{Q}=\int d{\bf r} \left (\Gamma_{Q}^{bulk}(r)  
\left( \frac{\epsilon_{F}(r)}{\epsilon_{F}(0)}\right)^{\beta+0.5}+
\Gamma_{Q}^{surf}(r) 
[1-\left(\frac {\epsilon_{F}(r)}{\epsilon_{F}(0)}\right)^{\beta}] \right) 
\equiv \int d{\bf r} \Gamma_{Q}(r)
\end{equation}
and
\begin{equation}
\Gamma_{D}=\int d{\bf r} \left (\Gamma_{D}^{bulk}(r)  
\left( \frac{\epsilon_{F}(r)}{\epsilon_{F}(0)}\right)^{\beta+1.5}+
\Gamma_{D}^{surf}(r) 
[1-\left(\frac {\epsilon_{F}(r)}{\epsilon_{F}(0)}\right)^{\beta}] \right) 
\equiv \int d{\bf r} \Gamma_{D}(r)
\end{equation}

We determine the nuclear
density in the Thomas-Fermi approximation using a Wood-Saxon potential with
a depth $V_0=-44$ MeV, thickness $a=0.67$ fm and sharp radius
$R_0= 1.27 A^{1/3}$ fm. We perform the calculations with 
the SkM force with parameters $\alpha=1/6$, $x_{0}=0.09$, $t_{0}=-2645 MeV fm^3$, 
$t_{1}=410 MeV fm^{5}$, $t_{2}=-135 MeV fm^{5}$, and $t_ {3}=15,595 MeV fm^{7/2}$.
For the mass dependence of the resonance energies for spherical medium mass and heavy nuclei we use 
the formulas,
$\hbar \omega= 64 A^{-1/3}$ MeV for GQR and $\hbar \omega=80 A^{-1/3}$ MeV for GDR.
Figures 1 and 2 illustrate the relative contribution of the damping widths of
GDR and GQR for a nucleus with $A=120$ as function of $r$. In these figures and also in the other 
figures, the dashed and dotted lines show the result of the calculations with the effective 
mass $m^*$ and the bare mass $m$, respectively. The sharp rise of  $\Gamma(r)$ in the vicinity of the 
surface is largely due to the effective mass, which is small in the bulk and approaches to its bare value at 
the surface, and to a lesser extend due to the increase of the Skyrme cross-section
at low densities. 
For comparison, the results for constant cross-sections of
$\sigma_0=30$ mb and $\sigma_{pn}=40$ mb are shown in the same figures by solid
lines. These constant cross-sections  correspond to a zero range force with a
strength $t_0=-300 MeV fm^3$  and all other parameters are set equal to zero
in (45). Figures 3 and 4 show the atomic mass dependence of
the GDR and GQR widths and comparison with data, respectively. The SkM force
with the effective mass underestimates the average trend of GDR for medium weight and heavy nuclei by 
about a factor two. The calculations with the bare mass gives a better description of the
average trend. In the GQR-case, the discrepancy between the calculations and
the average trend of data is larger than in the GDR-case. 
In figures 5 and 6, the measured GDR widths in $^{120}Sn $ and $^{208}Pb$ nuclei are plotted
as a function of temperature, and compared with the calculations performed with the
effective mass and the bare mass shown by dashed and dotted lines, respectively. The calculations
with the effective mass provide a reasonable description of the temperature dependence of data, but
the magnitude of damping is underestimated in both cases. The calculation with the bare mass give larger
damping, but the damping widths appear to grow faster than data as a function of temperature. 

\section{Conclusions}

In the standard nuclear transport models (mean-field transport models and their stochastic extensions in 
semi-classical or quantal form), the binary collisions are treated in a Markovian approximation by 
assuming the duration time of collision is much shorter 
than the mean-field fluctuations and the mean-free-time between collisions, 
which would be appropriate if two-body collisions can be considered 
instantaneous. As a result, the standard model provides a classical description
of transport properties of collective motion that is valid at low frequency-high temperature limit. When 
the system possesses fast collective modes, the standard description breaks down and it is necessary to 
incorporate memory effect associated with the finite duration
of binary collisions. This yields a non-Markovian extension of the transport description in which the basic 
transition rate is modified by involving  a direct coupling between collective modes and two-body 
collisions. The extended model leads to a description of
the transport properties of collective modes that is in accordance with the quantal fluctuation-dissipation 
relation. In this work, we present a detailed derivation of the collisional widths of  isoscalar and isovector  
collective nuclear vibrations in both quantal
and semi-classical frameworks by considering the linearized limits of the extended TDHF and the BUU 
model with non-Markovian collision term. The standard treatment with a Markovian collision term leads 
to vanishing collisional widths at zero temperature. Whereas in the non-Markovian treatment the 
collisional widths are finite and consistent with the Landau's expression for damping of zero sound in 
Fermi liquids. The numerical result of the collisional damping is rather sensitive to the in-medium
nucleon-nucleon cross-sections around Fermi energy, for which an accurate information is not available.
In the present investigations, by employing an effective Skyrme force with SkM parameters, 
we carry out calculations of  the damping widths of  giant quadrupole and giant dipole excitations
in semi-classical framework, and compare the results with the GDR measurements
in $^{120}Sn$ and $^{208}Pb$ nuclei at finite temperatures. In particular for GDR, the magnitude of the 
collisional damping with the bare nucleon mass is a sizable fraction of the observed damping widths at 
zero temperature, however the effective mass further reduces the magnitude of damping in both cases. 
Aside from the
magnitude, calculations are qualitatively in agreement with the broadening of GDR widths as a
function of temperature in both $^{120}Sn$ and $^{208}Pb$ nuclei. 

One of the main aims of the present investigation was to assess how much of the
the total width of GRE's is exhausted by decay into the incoherent 2p-2h
states. The calculations have been performed within the Thomas-Fermi
approximation, which is known from independent studies to be very reliable
for description of the 2p-2h level densities \cite{SchHas}. However,
our results remain semi-quantitative, since in the Thomas-Fermi framework
we need the in-medium cross-sections locally down to very low densities, i.e.
we need cross-sections which interpolate correctly between the free space and
the medium. At the moment such cross-sections are not available (at least
not analytically), and thus we were forced to invent our own interpolation
scheme, which, although reasonable, is subject to some uncertainties.
We found that a sizeable fraction of $\Gamma^{\downarrow}$ is accounted 
for by the
incoherent decay. This is the case, for instance, for the GDR and also to a
lesser extend for the GQR. In addition, we found for the GDR that the
percentage of the incoherent decay, depending somewhat on the nucleon 
effective mass, strongly increases with tempearture. This finding is
not very surprizing, since at temperature $T>3$ MeV shell effects are
absent and the collectivity of the vibrational states is strongly reduced.
Therefore to a good approximation, a hot nucleus can be regarded as a 
finite blob of a hot Fermi gas. Inspite of this, in particular at lower
temperatures, the influence of the low-lying collective sates on the
damping is missing in our description. Also, the question of the saturation
of the GDR width has not been adressed in this work. Further studies are 
needed
for a quantitative  description of the damping of nuclear Giant Resonances. 
 
\begin{center}
{\bf Acknowledgments}
\end{center}

One of us (O.Y.) gratefully acknowledges Tennessee Technological University for support and warm 
hospitality extended to him during his visit. One of us (S. A.) likes to thank the  Physics Department of 
Middle East Technical University for warm hospitality extended to him during his visits. This work is 
supported in part by the U.S. DOE Grant No. 
DE-FG05-89ER40530.


\newpage

{\bf Figure Captions:}

\begin{description}

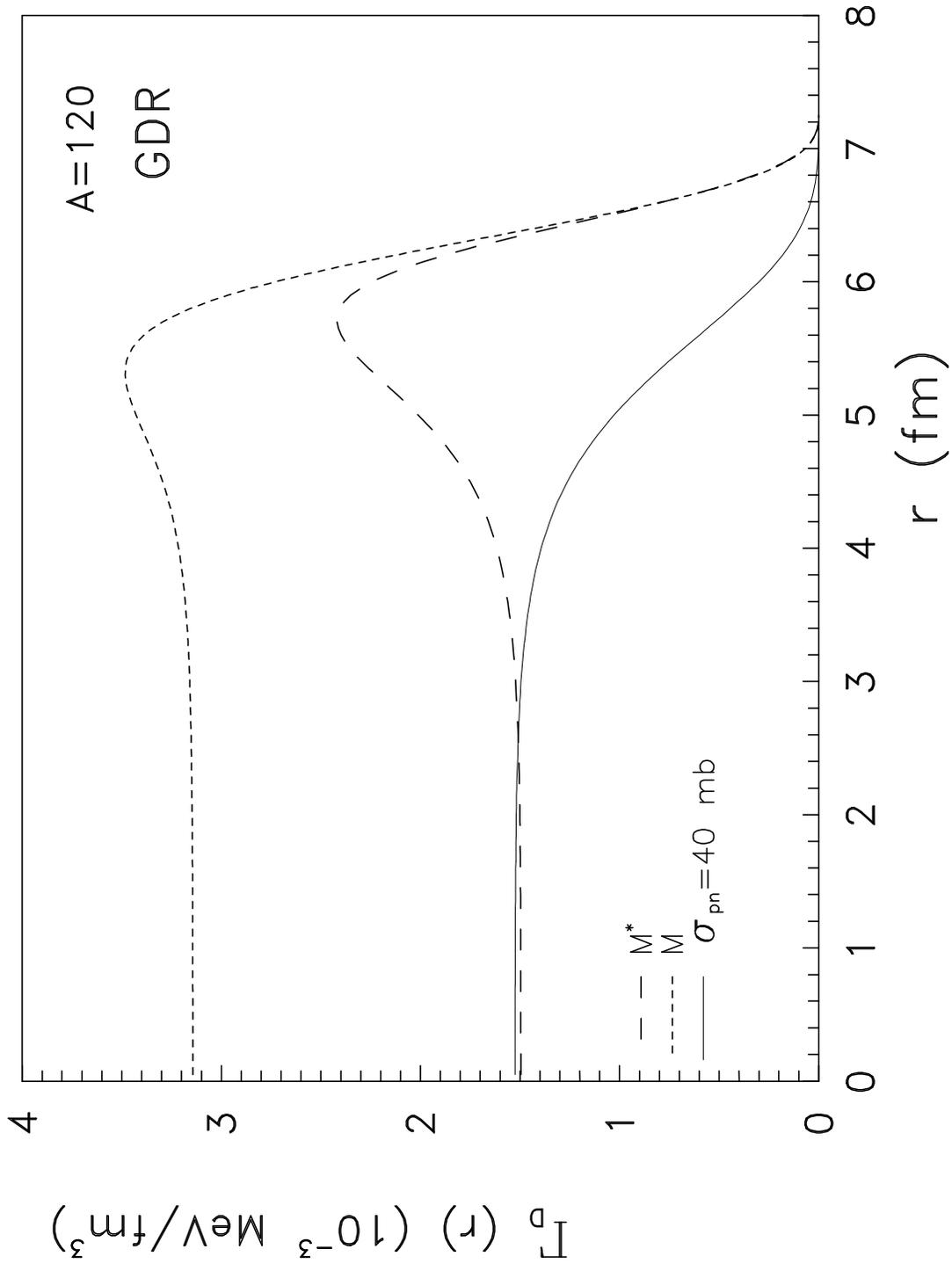
\item[{\bf Figure 1}:] The relative contribution to the collisional damping width of GDR as a function of 
$r$ for $A=120$ at zero temperature. Solid, dashed and dotted lines are
calculations with a constant cross-section $\sigma_{pn}=40$ mb, with the SkM force with
the effective mass and with the SkM force with the bare mass, respectively.

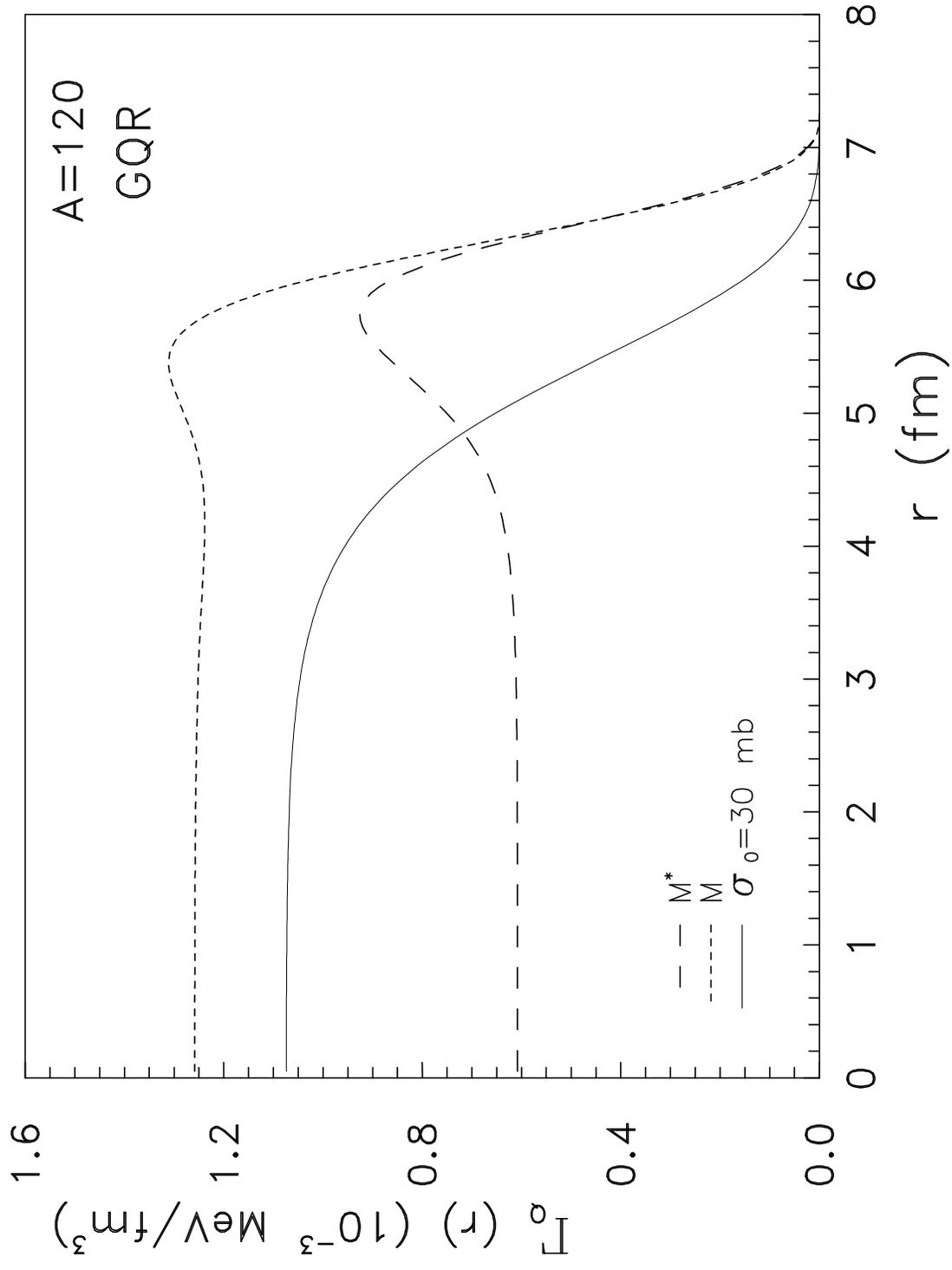
\item[{\bf Figure 2}:] The relative contribution to the collisional damping width of GQR as a function of 
$r$ for $A=120$ at zero temperature. Solid, dashed and dotted lines are
calculations with a constant cross-section $\sigma_{pn}=30$ mb, with the SkM force with
the effective mass and with the SkM force with the bare mass, respectively.

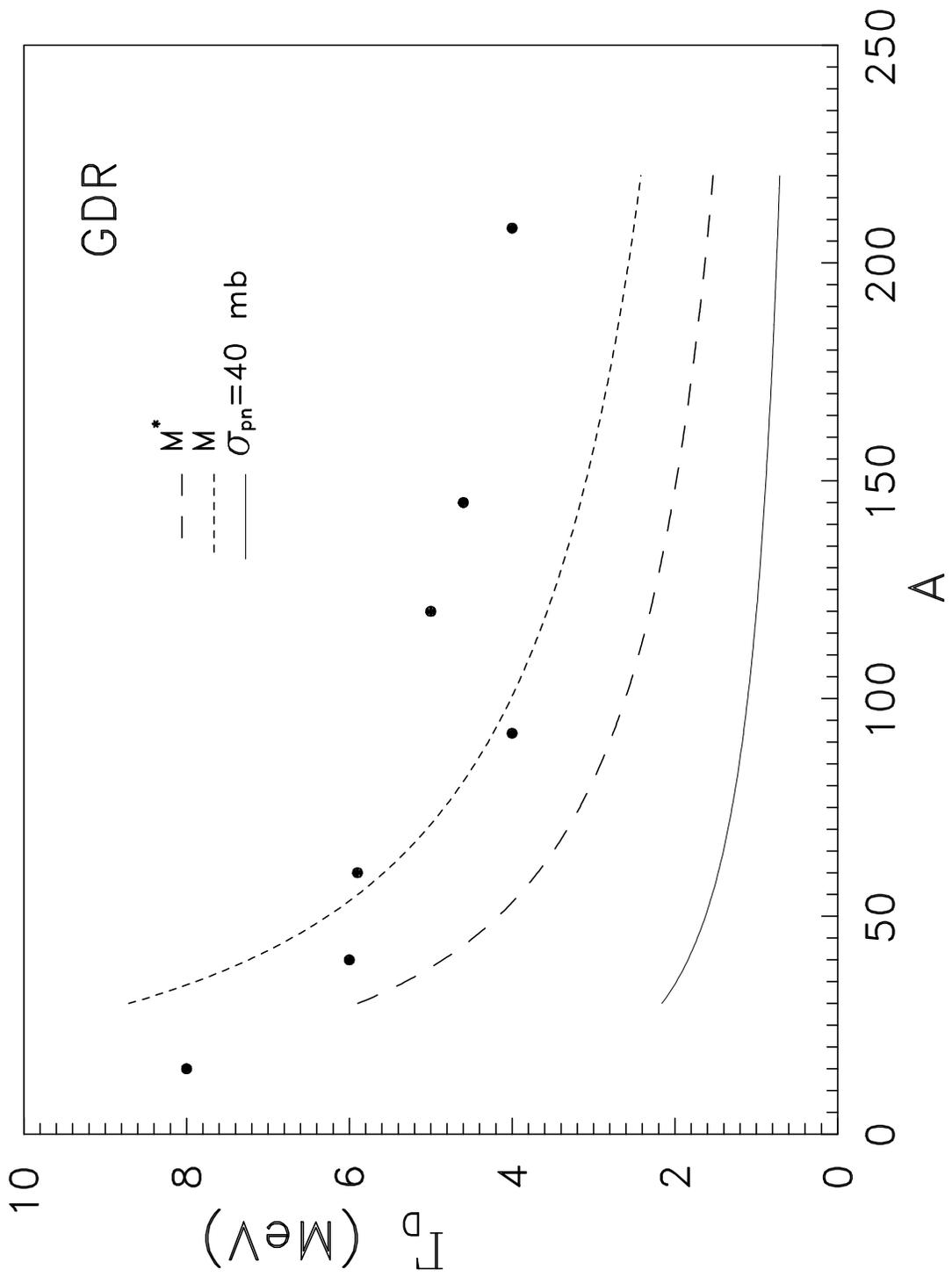
\item[{\bf Figure 3}:] The collisional damping width of GDR as a function of mass number $A$ at zero 
temperature. Solid, dashed, dotted lines  are calculations with a 
constant cross-section $\sigma_{pn}=40$ mb, with the SkM force with the 
effective mass, with the SkM force with the bare 
mass, respectively, and the points show the data.

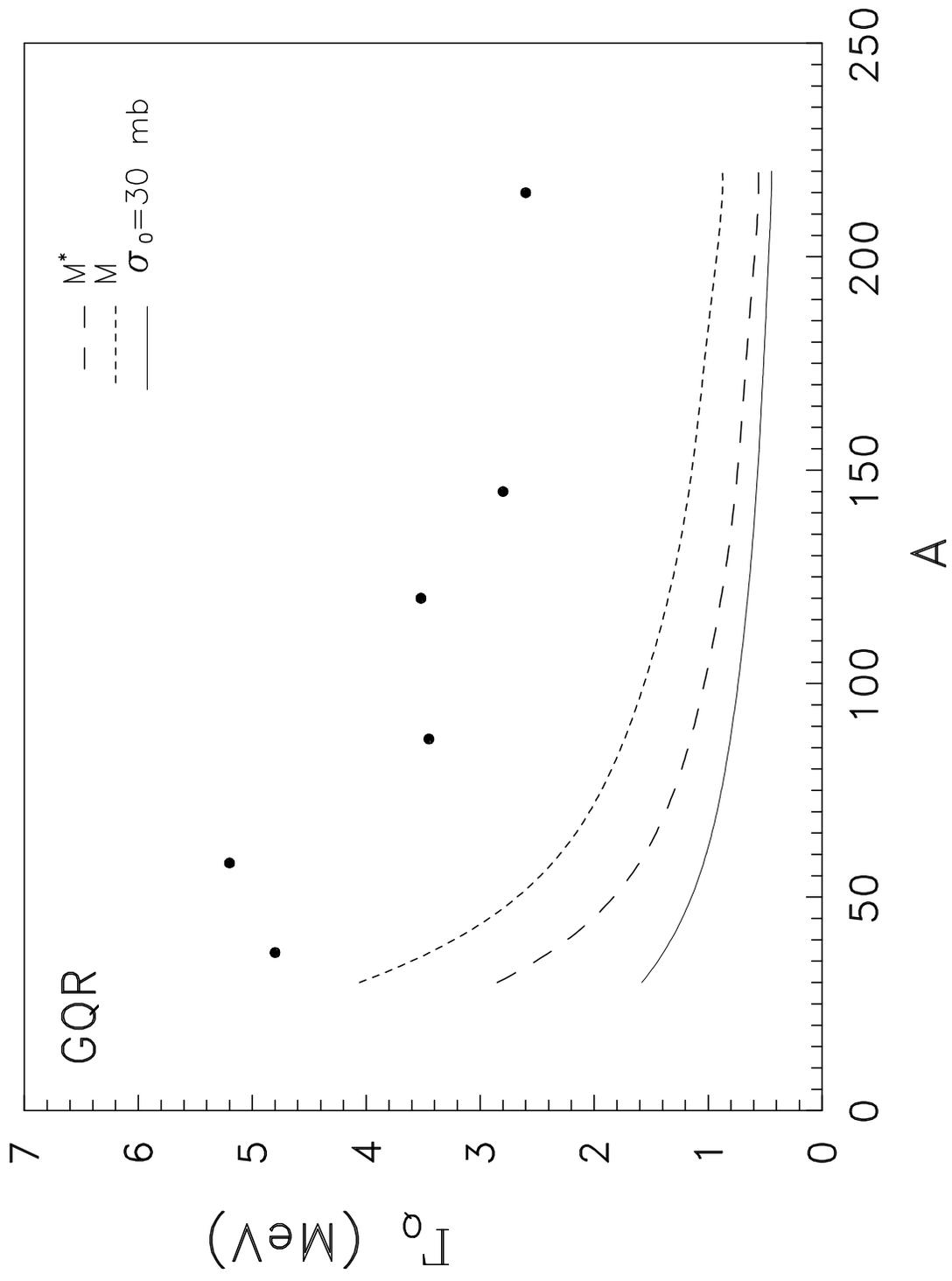
\item[{\bf Figure 4}:] The collisional damping width of GQR as a function of mass number $A$ at zero 
temperature. Solid, dashed, dotted lines  are calculations with a 
constant cross-section  $\sigma_{pn}=30$ mb, with the SkM force with the 
effective mass, with the SkM force with the bare 
mass, respectively, and the points show the data.

\item[{\bf Figure 5}:] The collisional damping width of GDR in $^{120}Sn $ as a function of 
temperature. Dashed, dotted lines and points are calculations with the SkM force with the effective mass, 
with the SkM force with the bare mass and data taken from \cite{Tin}, respectively.

\item[{\bf Figure 6}:] The collisional damping width of GDR in $^{208}Pb $ as a function of 
temperature. Dashed, dotted lines and points are calculations with the SkM force with the effective mass, 
with the SkM force with the bare mass and data taken from \cite{Lead}, respectively.

\end{description}

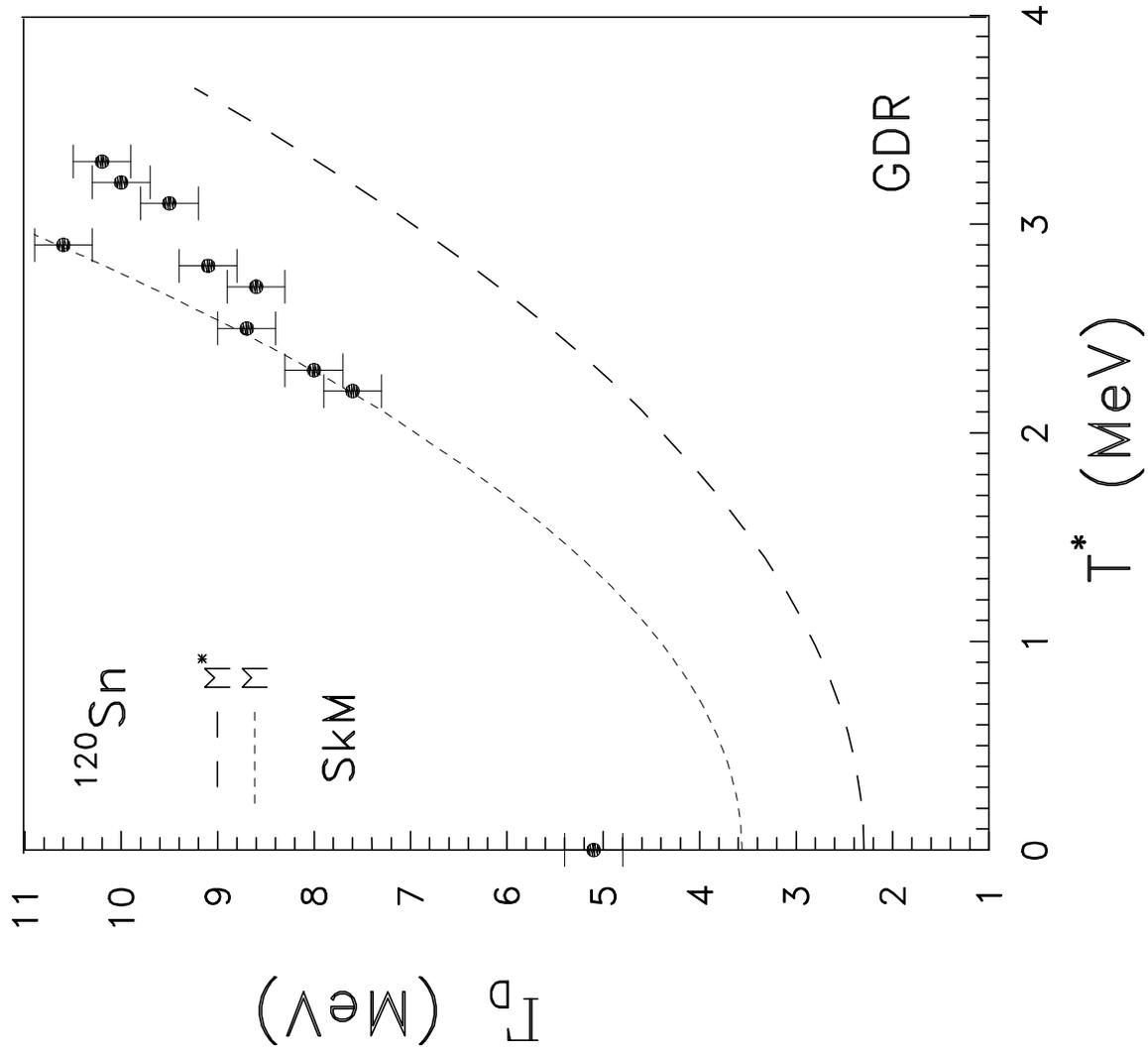
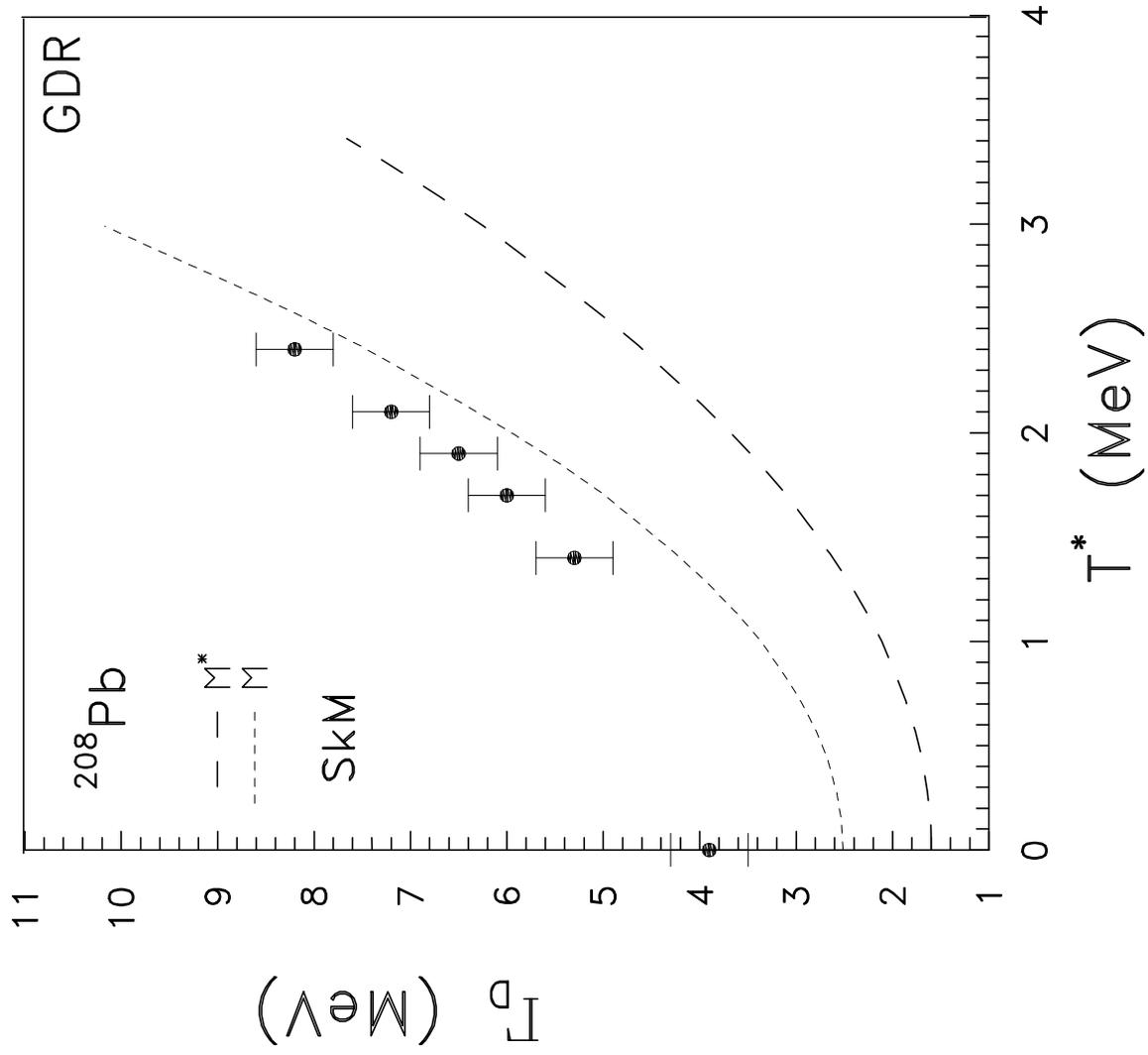
\end{document}